\documentclass[aps,prb,twocolumn,superscriptaddress]{revtex4-1}

\usepackage{graphicx}
\usepackage{dcolumn}
\usepackage{bm}

\newcommand{\chem}[1]{\ensuremath{\mathrm{#1}}}

\begin{document}

\title{Molecular Dynamics Simulation of Coherent Interfaces in Fluorite Heterostructures}

\author{Benjamin J. Morgan}
\affiliation{Department of Materials, University of Oxford, Parks Road, OX1 3PH, UK}
\affiliation{Stephenson Institute for Renewable Energy, Department of Chemistry, University of Liverpool, Liverpool, L69 3BX,UK}
\email{bmorgan@liv.ac.uk}
\author{Paul A. Madden}
\affiliation{Department of Materials, University of Oxford, Parks Road, OX1 3PH, UK}
\date{\today}

\begin{abstract}
The standard model of enhanced ionic conductivities in solid electrolyte heterostructures follows from a continuum mean-field description of defect distributions that makes no reference to crystalline structure. To examine ionic transport and defect distributions while explicitly accounting for ion--ion correlations and lattice effects, we have performed molecular dynamics simulations of a model coherent fluorite heterostructure without any extrinsic defects, with a difference in standard chemical potentials of mobile fluoride ions between phases induced by an external potential. Increasing the offset in fluoride ion standard chemical potentials across the internal interfaces decreases the activation energies for ionic conductivity and diffusion and strongly enhances fluoride ion mobilities and defect concentrations near the heterostructure interfaces. Non-charge-neutral ``space-charge'' regions, however, extend only a few atomic spacings from the interface, suggesting a continuum model may be inappropriate. Defect distributions are qualitatively inconsistent with the predictions of the continuum mean-field model, and indicate strong lattice-mediated defect--defect interactions. We identify an atomic-scale ``Frenkel polarisation'' mechanism for the interfacial enhancement in ionic mobility, where preferentially oriented associated Frenkel pairs form at the interface and promote local ion mobility via concerted diffusion processes.
\end{abstract}
\maketitle

\section{Introduction}
\begin{figure*}[htb]
  \begin{center}
    \resizebox{14cm}{!}{\includegraphics*{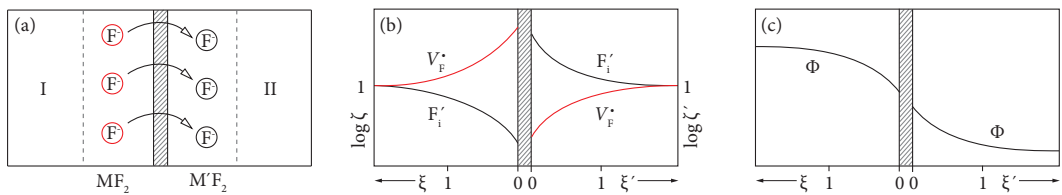}} %
    \caption{\label{fig:Space_Charge_Schematic} Schematic of the ion redistribution predicted by the Space-Charge Model at an heterointerface between two Frenkel disordered \chem{MF_2} and M$^\prime$F$_2$ phases. (a) Mobile F$^-$ ions have a lower standard chemical potential in phase II, and move across the interface to form interstitials, leaving behind vacancies. (b) The electrostatic interaction between these excess defect populations constrains the space charges to the near-interface region, producing characteristic  defect concentration profiles. (c) An excess of vacancies or intersitials corresponds to local non-charge-neutrality and curvature of the mean electrostatic potential, $\Phi$.\cite{Maier_BerBunsengesPhysChem1985, Maier_ProgSolidStatChem1995, Maier_JPhysChemSol1985, Maier_JElectrochemSoc1987}  $\zeta=c/c_\infty$, where $c_\infty$ is the bulk concentration of each defect species. $\xi=x/\lambda_\mathrm{D}$, where $\lambda_\mathrm{D}$ is the Debye length.}
  \end{center}
\end{figure*}

Heterostructured solid electrolytes can have ionic conductivities orders of magnitude greater than their component bulk phases, making these materials candidates for high performance electrolytes in devices such as batteries and fuel cells.\cite{GuoAndMaier_AdvMater2009, SataEtAl_Nature2000} This behaviour has been explained generically by noting that in a heterostucture the standard chemical potential of mobile ions can be expected to differ across the heterointerface, which provides a driving force for these ions to spontaneously redistribute across the interface into locally non-stoichiometric ``space-charge'' regions (Fig.~\ref{fig:Space_Charge_Schematic}a).\cite{Maier_BerBunsengesPhysChem1985, Maier_ProgSolidStatChem1995, Maier_JPhysChemSol1985, Maier_JElectrochemSoc1987} The mutual electrostatic interaction between these space charges confines them close to the interface, where locally increased defect concentrations greatly enhance net ionic conductivities. 

This schema of defect redistribution into space-charge regions emerges from solving the Poisson-Boltzmann equation within a continuum mean-field description: component heterolayers are considered as continuum dielectrics, and any ionic defects are considered as point charges that interact solely through a mean-field Coulomb term. For heterostructures with defect chemistries characterised by intrinsic Frenkel disorder the predicted equilibrium distribution of ions consists of an excess of mobile ions in one phase, equivalent to an increase in interstitials, and a deficiency of mobile ions in the second phase, equivalent to an increase in vacancies, and with opposite formal charge (Fig.~\ref{fig:Space_Charge_Schematic}b). This defect excess corresponds to local non-charge-neutrality and an associated curvature of the mean electrostatic potential profile that characterise space-charge regions (Fig.~\ref{fig:Space_Charge_Schematic}c).

In contrast with the underlying approximations of a continuum mean-field treatment, the distributions of ions and defects in crystalline electrolytes are strongly inhomogeneous. In a perfect crystal ions occupy well-defined lattice sites. Point defects are defined as local deviations from the perfect lattice, with their positions constrained by the lattice structure: interstitials occupy interstitial sites, and vacancies exist as absences within the periodic host lattice. The lattice structures of component heterolayers can therefore be expected to modulate the equilibrium distribution of mobile ions, while strain effects within a crystalline lattice can mediate non-Coulombic interactions between defects.\cite{MorganAndMadden_PhysRevLett2011} It is not known, however, to what extent the predictions of the continuum mean-field model remain appropriate where crystal structure is taken into consideration.

To examine the relationships between crystal structure, equilibrium defect distribution, and ionic mobility, we have performed atomistic molecular dynamics simulations of a model fluorite heterostructure under intrinsic conditions. At a mesoscopic scale, our model heterostructure is analogous to the Frenkel-disordered heterointerface, described by Maier in Ref.~\onlinecite{Maier_BerBunsengesPhysChem1985}. Our atomistic model, however, describes electrostatic and excluded volume correlations between mobile ions and the interactions between mobile ions and the host lattice, in contrast to previous continuum mean-field analyses. Focussing on intrinsic defect conditions allows us to make a direct comparison with the corresponding continuum mean-field Gouy-Chapman model.\cite{Maier_BerBunsengesPhysChem1985, Maier_ProgSolidStatChem1995} In heterostructured systems where extrinsic dopants contribute significantly to the defect chemistry Mott-Schottky (extrinsic) conditions should be considered, and behaviour may differ strongly from that of an otherwise equivalent system under Gouy-Chapman conditions.\cite{Jamnik_SolStatIonics2006, GuoEtAl_PhysRevB2007, GuoAndMaier_AdvFunctMater2009} It is therefore not possible to directly compare the results of our simulations with experimental data for aliovalently doped heterostructures, although the questions of lattice structure effects remain pertinent in these materials.

Previous molecular dynamics simulations of ionic transport in heterostructures have focused on describing specific materials; in particular heterostructured \chem{CaF_2}/\chem{BaF_2},\cite{NomuraAndKobayashi_Ionics2003, SayleEtAl_ChemComm2003, SayleEtAl_PhysChemChemPhys2005b, AdamsAndTan_SolStatIonics2008, ZahnEtAl_JPhysChemC2009} which is considered a model system for understanding interface effects on ionic conductivities. The bulk lattice parameters of \chem{CaF_2} and \chem{BaF_2} differ by $14\%$, and the interfaces within the heterostructures  are highly strained. In experimental samples this strain is relieved by misfit dislocations that form at the interface,\cite{Jin-PhillippEtAl_JChemPhys2004} and equivalent dislocations can be introduced by simulated annealing in \chem{CaF_2}/\chem{BaF_2} simulations.\cite{ZahnEtAl_JPhysChemC2009} Interfacial strain can strongly affect ionic conductivities of lattice-mismatched heterostructure interfaces, by modifying local defect numbers or producing low energy diffusion pathways that allow rapid transport of ions.\cite{PennycookEtAl_PhysRevLett2010, FabbriEtAl_SciTechnolAdvMat2010, Rupp_SolStatIonics2012} These effects can obscure the role of space-charge formation in ionic conductivity enhancement, and interfere with the analysis of this mechanism when simulating lattice mismatched heterostructures.\cite{ZahnEtAl_JPhysChemC2009, AdamsAndTan_SolStatIonics2008, SayleEtAl_ChemComm2003}

\begin{figure}[tb]
  \begin{center}
    \resizebox{8.2cm}{!}{\includegraphics*{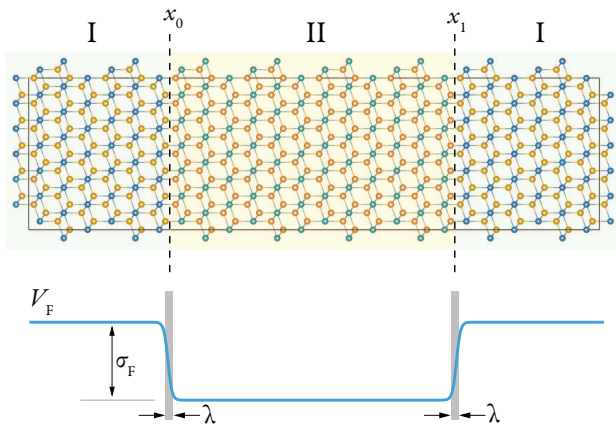}} %
    \caption{\label{fig:simulation_construction} (Color online) (upper panel) Snapshot of the simulation cell, indicating regions I and II. (lower panel) Variation in $V_\mathrm{F}$ across the simulation cell, indicating the parameters for Eqn.\ \ref{eqn:external_potential}.}
  \end{center}
\end{figure}

\section{Simulation Methods}
In the continuum mean-field model, the variation in chemical environment across a heterostructure interface is described only as a difference in the standard chemical potential of mobile ions in each heterolayer.\cite{Maier_BerBunsengesPhysChem1985, Maier_ProgSolidStatChem1995, Maier_JPhysChemSol1985, Maier_JElectrochemSoc1987} We have considered a closely analogous atomistic system of a symmetric \chem{CaF_2}/\chem{CaF_2} heterostructure, with an offset in the standard chemical potential for the mobile $\mathrm{F}^-$ ions described by an external potential. This construction allows the role of lattice structure to be examined in an otherwise ideal model system, where the difference in standard chemical potential of mobile ions between heterolayers can be controlled independently and the misfit strain inherent to conventional heterostructure simulations is minimised. 

\begin{figure}[tb]
  \begin{center}
    \resizebox{6.0cm}{!}{\includegraphics*{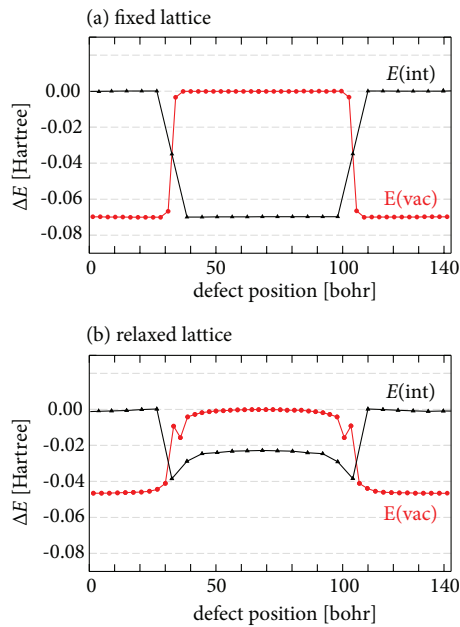}} %
    \caption{\label{fig:CaF2_defect_energies}(Color online) Variation in total energies for single vacancies (red circles) and interstitials (black triangles) as a function of $x$. Calculations were performed on perfect initial structures, with individual vacancies or interstitials introduced at all possible inequivalent positions in the cell. (a) All ions are held fixed at lattice or ideal defect sites. (b) All ion positions are allowed to fully relax.}
  \end{center}
\end{figure}

Our simulation model consists of a coherent \chem{CaF_2} ``homostructure'' of 864 molecular units, in a periodic $142.8\times43.7\times37.9\,\mathrm{bohr}$ cell ($1\,\mathrm{bohr} = 0.0529\,\mathrm{nm}$, which is the zero-pressure volume at $0\,\mathrm{K}$ with the crystal oriented with the $\left[111\right]$ direction along $x$ (Fig.\ \ref{fig:simulation_construction}). Interatomic interactions are described using the rigid-ion potential of Gillan,\cite{Gillan_JPhysC1986a} which was constructed to reproduce the experimental zero-temperature lattice parameter and low-temperature Frenkel pair formation energy, and gives a good account of the superionic transition. To model an offset in standard chemical potential we include an external potential, $V_\mathrm{F}(x)$, that acts only on the F$^-$ ions: 
\begin{equation}
  \label{eqn:external_potential}
  V_\mathrm{F} = \sigma_\mathrm{F}\left[ \left(1+\mathrm{e}^\frac{x-x_0}{\lambda}\right)^{-1} + \left(1+\mathrm{e}^\frac{-(x-x_1)}{\lambda}\right)^{-1} \right].
\end{equation}
The external potential defines a pair of antisymmetric interfaces, centered at $x_0$ and $x_1$ and each with a width of $\lambda$. Providing $\lambda$ is small with respect to the distance between $x_0$ and $x_1$ the regions away from the interfaces can be considered bulk-like with locally-constant (but different) F$^-$ standard chemical potentials. For all calculations, fixed parameters for the external potential were $\lambda=0.5\,\mathrm{bohr}$ and $x_0=32.6\,\mathrm{bohr}$, $x_1=104.0\,\mathrm{bohr}$, with the potential strength, $\sigma_\mathrm{F}$, varied between groups of simulations. If lattice relaxation effects are neglected, as in the continuum model, the standard chemical potentials of the F$^-$ ions in the bulk regions are simply offset by $\sigma_\mathrm{F}$, and the standard chemical potentials of interstitial and vacancy defects are also shifted by $\pm\sigma_\mathrm{F}$ (Fig.\ \ref{fig:CaF2_defect_energies}a).\footnote{Within the Ewald sum the $k=0$ term is undefined, making such a calculation equivalent to calculating the energy for the periodic charged system in the presence of a uniform charge-compensating background. Energies for systems with different charges cannot be directly compared, but energy \emph{differences} for calculations with the same charge and cell dimensions are independent of any charge corrections and therefore meaningful.}

In a dynamic atomistic simulation, as in a real crystal, defect chemical potentials and formation energies depend on lattice relaxation. When ion positions are relaxed the defect formation energies are no longer simply related to $V_\mathrm{F}\left(x\right)$ (Fig.\ \ref{fig:CaF2_defect_energies}b).
Because the external potential acts on all F$^-$ ions the lattice relaxation associated with defect formation depends on the defect position.
In the bulk regions away from the interface defect formation energies are approximately constant, but the offset between the two regions is reduced compared to the fixed lattice case. This is a consequence of the relaxation of the lattice in response to the applied external potential, which expands region I along $x$, destabilising vacancies and stabilising interstitials relative to the bulk lattice, and compresses region II along $x$, stabilising vacancies and destabilising intersitials relative to the bulk lattice. This lattice strain along $x$ is an artifact of our approach to realise the continuum model in an atomistic molecular dynamics simulation. In addition, close to the interfaces the local lattice relaxation is asymmetric and defect formation energies do not vary linearly with $V_\mathrm{F}$.
The predominant effect of the applied potential, however, is that for $\sigma_\mathrm{F}\ge0.04\,\mathrm{Hartree}$ ($1\,\mathrm{Hartree}=27.211\,\mathrm{eV})$ isolated vacancies are stabilised in region I relative to region II, and isolated interstitials are stabilised in region II relative to region I.\footnote{See Supplemental Material at [URL will be inserted by publisher] for plots of relative energies of isolated vacancies and interstitials as a function of $\sigma_\mathrm{F}$.} For defects interacting only via a mean-field electrostatic term, this would be expected to produce a strong driving force for interstitials and vacancies to segregate across the interface to form space charges, as in the continuum model.

We performed constant volume molecular dynamics simulations of $500\,000$ steps of timestep $100\,\mathrm{au}$ ($1\,\mathrm{au}=2.419\times10^{-17}\,\mathrm{s}$; simulation lengths $\approx1\,\mathrm{ns}$) for a range of temperatures at each value of $\sigma_\mathrm{F}$, which was varied from $\sigma_\mathrm{F}=0.00$ to $0.08\,\mathrm{Hartree}$. All simulations were performed below the experimental superionic transition temperature of $1420\,\mathrm{K}$.\cite{DworkingAndBredig_JPhysChem1968} Simulation temperatures were controlled by a Nos\'{e}-Hoover thermostat with a relaxation time of $20000\,\mathrm{au}$. 

\section{Results}
\subsection{Ionic transport coefficients}
\begin{figure}[tb]
  \begin{center}
    \resizebox{8cm}{!}{\includegraphics*{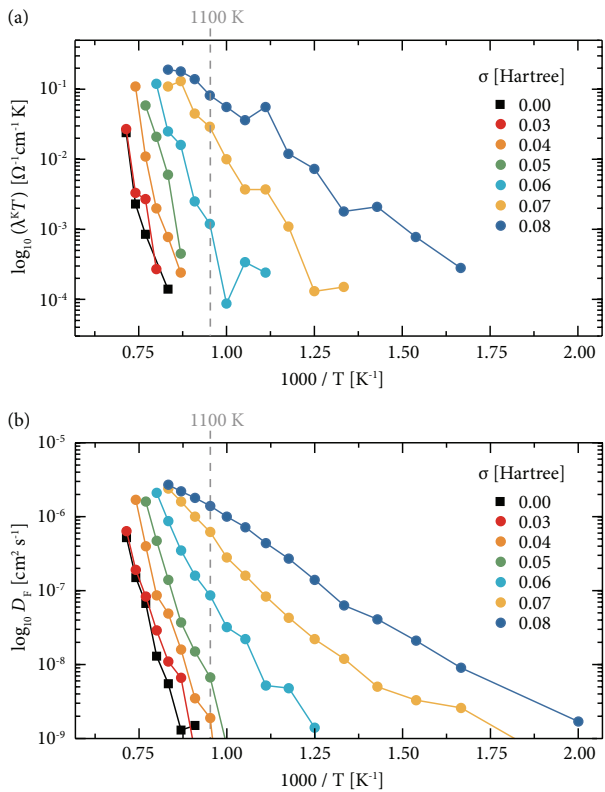}} %
    \caption{\label{fig:CaF2_heterostructure_diffusion}(Color online) (a) Calculated conductivities, $\lambda^\mathrm{K}$, and (b) diffusion coefficients, $D_\mathrm{F}$, for $\sigma_\mathrm{F}=0.00, 0.03$--$0.08$. The dashed line highlights data points at $T=1100\,\mathrm{K}$.}
  \end{center}
\end{figure}

Ionic conductivities were calculated from the simulation trajectories as the long-time limit of the mean-squared displacement of the charge density,\cite{MorganAndMadden_JChemPhys2004} 
\begin{equation}
  \lambda^\mathrm{K}=\frac{\beta e^2}{V}\lim_{t\to\infty}\left(6t\right)^{-1}\left<\left|Q_+\Delta_+(t)+Q_-\Delta_-(t)\right|^2\right>,
\end{equation}
where $\Delta_\alpha(t)$ is the net displacement of all the ions of species $\alpha$ in time $t$,
\begin{equation}
  \Delta_\alpha(t) = \sum_{i\in\alpha}\delta\vec{r}_i(t),
\end{equation}
and $\delta\vec{r}_i(t)$ is the displacement of ion $i$ in time $t$.
Fluoride ion diffusion coefficients, $D_\mathrm{F}$, were calculated using the Einstein relation
\begin{equation}
  D_{F} = \lim_{t\to\infty}\frac{\left< \left| \delta \vec{r}_\mathrm{F}(t)\right|^2 \right> }{6t},
  \label{eqn:diffusion_coefficient}
\end{equation}
where $\delta\vec{r}_\mathrm{F}(t)$ is the displacement of a fluoride ion in time $t$. Arrhenius plots of $\log_{10} D_\mathrm{F}$ versus $1000/T$ for $\sigma_\mathrm{F}=0.00$ and $0.03$--$0.08\,\mathrm{Hartree}$ are shown in Fig.\ \ref{fig:CaF2_heterostructure_diffusion}. For $\sigma_\mathrm{F}\le0.03$ there is only a small increase in diffusion coefficients with increasing $\sigma_\mathrm{F}$. At $\sigma_\mathrm{F}=0.03$ the increase  in diffusion coefficients relative to the bulk $\sigma_\mathrm{F}=0.00$ data is $\sim50\%$ across this temperature range. This effect can be attributed to the strain along $x$ produced by the external potential, which allows Frenkel pairs to form slightly more easily in region I.  For $\sigma_\mathrm{F}\ge0.04$, the diffusion coefficient increases much more rapidly relative to the bulk values, and for $\sigma_\mathrm{F}\ge0.06$ the activation energy also decreases. The much stronger enhancement for $\sigma_\mathrm{F}\ge0.04$ can be attributed to the external potential now stabilising independent vacancies in region I and independent interstitials in region II. The change in slope of these Arrhenius plots means that the differences in behaviour for different values of $\sigma_\mathrm{F}$ are magnified at lower temperatures. In the following analysis we focus on data for $T=1100\,\mathrm{K}$, which is the lowest temperature that gives good statistics for the diffusion coefficient for all values of $\sigma_\mathrm{F}\ge0.04$, and is considerably lower than the superionic transition temperature predicted by the Gillan potential of $\approx1400\,\mathrm{K}$.\cite{Gillan_JPhysC1986a} The temperature of $1100\,\mathrm{K}$ was chosen for convenience, but the Arrhenius plots in Fig.\ \ref{fig:CaF2_heterostructure_diffusion} do not exhibit large changes in slope as the temperature is decreased, which indicates that the same transport mechanisms are operative at lower temperatures. 

\begin{figure*}[tb]
  \begin{center}
    \resizebox{18cm}{!}{\includegraphics*{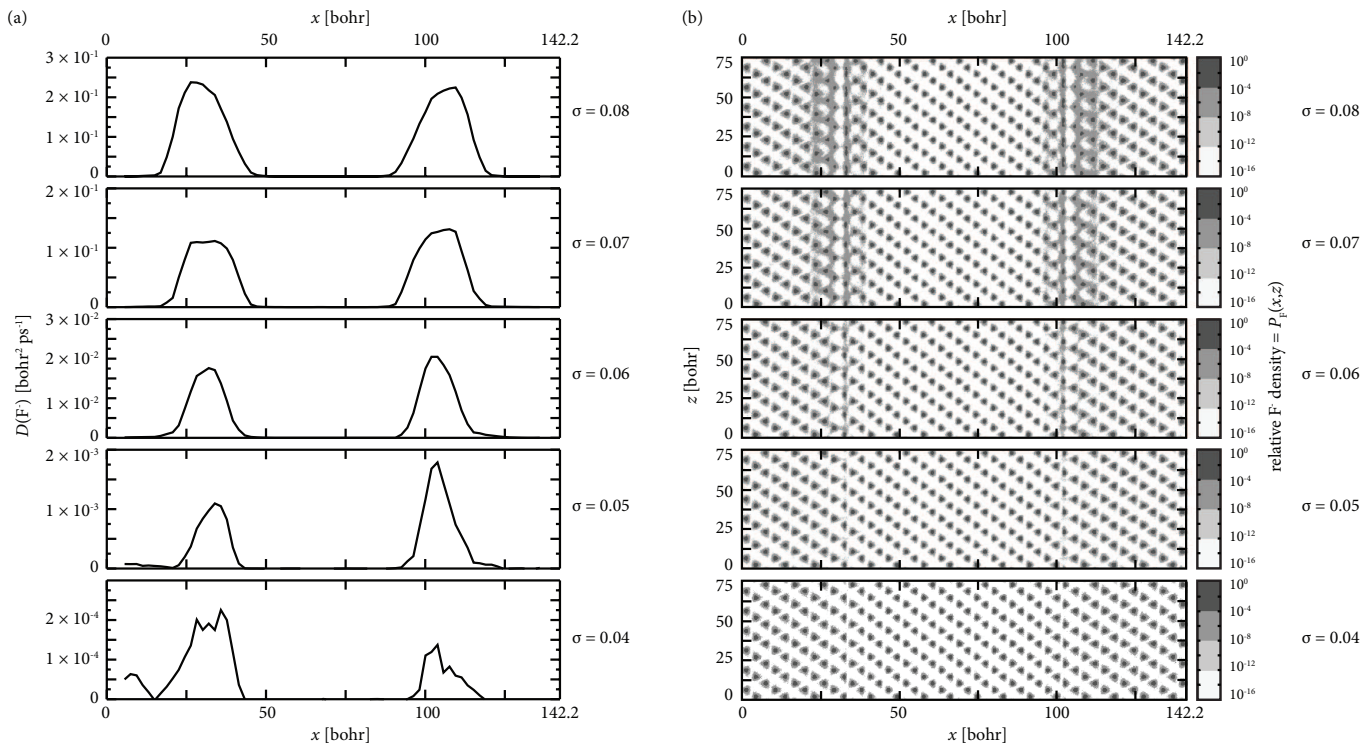}} %
    \caption{\label{fig:CaF2_heterostructure_diffusion_profile}(a) Variation in F$^-$ ion diffusion coefficients with $x$-position in the simulation cell, as a function of $\sigma_\mathrm{F}$. (b) Time-averaged fluoride ion density in the $xz$ plane. The density scale is normalised to give the two-dimensional probability density of finding a fluoride ion with a given $x$ and $y$ at any step of the simulation.}
  \end{center}
\end{figure*}

Local $\mathrm{F}^-$ diffusion coefficients as a function of $x$ are obtained by averaging $D_\mathrm{F}$ over all fluoride ions present at time $t$ in $yz$-layers of the simulation cell. These local diffusion coefficient profiles for $\sigma_\mathrm{F}=0.04$--$0.08$ and $T=1100\,\mathrm{K}$ are shown in Fig.\ \ref{fig:CaF2_heterostructure_diffusion_profile}a. Diffusion rates are strongly enhanced near the interfaces, and this local enhancement increases over three orders of magnitude as $\sigma_\mathrm{F}$ increases from $0.04$ to $0.08$. Increasing $\sigma_\mathrm{F}$ also increases the width of the region of enhanced diffusion coefficients. 

Calculated local average $\mathrm{F}^-$ ion densities---plotted in Fig.\ \ref{fig:CaF2_heterostructure_diffusion_profile}b for $\sigma_\mathrm{F}=0.04$--$0.08$ and $T=1100\,\mathrm{K}$, and normalised to give two-dimensional probability distributions for finding a fluoride ion at specific $x$ and $z$ coordinates---confirm the F$^-$ mobility enhancement close to the interface upon increasing $\sigma_\mathrm{F}$ values.  At low values of $\sigma_\mathrm{F}$ the fluoride ions principally vibrate around their ideal lattice positions, with the tetrahedral coordination environment producing a characteristic triangular motif in this cell orientation. For transport to occur, ions must, at least momentarily, move away from their lattice sites. At low $\sigma_\mathrm{F}$ values the probability of finding a fluoride ion away from a lattice site is small, corresponding to low $\mathrm{F}^-$ transport. As $\sigma_\mathrm{F}$ increases the fluoride ion densities close to the interface positions are ``smeared out''---$\mathrm{F}^-$ ions are much more likely to move away from their ideal lattice sites, consistent with the enhanced F$^-$ diffusion in these regions observed in the local diffusion profiles in Fig.\ \ref{fig:CaF2_heterostructure_diffusion_profile}a. The contiguous intersite $\mathrm{F}^-$ density at high $\sigma_\mathrm{F}$ values confirms enhanced diffusion of the fluoride ions parallel to the $yz$ plane of the interface.

\subsection{Defect distributions and mean electrostatic potentials}
Within the continuum mean-field model the enhanced conductivities observed in ionically conducting heterostructures are explained as a consequence of ionic redistribution near the interface into space charge regions. To be precise, under intrinsic (Gouy-Chapman) conditions an excess of mobile ions is expected where the ion standard chemical potentials are low (region II in our model system), and a corresponding deficiency of these ions will be found on the side where their standard chemical potentials are high (region I).\cite{Maier_BerBunsengesPhysChem1985, Maier_ProgSolidStatChem1995} Expressing this as populations of interstitial and vacancy defects, we would expect formation of paired space charge near each interface, with excess interstitials in region II and excess vacancies in region I (c.f.\ Fig.\ \ref{fig:Space_Charge_Schematic}b). Importantly, in each space charge region where the concentration of one defect is enhanced, the concentration of the counter defect is predicted to be \emph{decreased}. More generally, the one-dimensional Poisson-Boltzmann equation used in the continuum mean-field model directly couples the local concentration of each defect species to the local electrostatic potential, and has no solutions where concentrations of oppositely charged defects are simultaneously enhanced at any single position.

In \chem{CaF_2} the calcium ions vibrate about fixed lattice positions, and their instantaneous positions define a space-filling set of tetrahedra and octahedra that can be occupied by the mobile F$^-$ ions. At any point in the simulation trajectory the instantaneous ion positions can be used to determine whether each F$^-$ ion occupies either a lattice tetrahedron or an interstitial octahedral site. The same analysis assigns instantaneous vacancy positions to the centres of unoccupied  tetrahedra. At the temperatures of interest the fluoride ions may perform large amplitude vibrations about their average sites without contributing to F$^-$ diffusion. When calculating defect populations we therefore consider only ``significant'' defects, defined as sites where the assignment of a specific ion is unchanged for a minimum of two consecutive analysis frames ($0.48\,\mathrm{ps}$). This follows the approach used in previous studies of ionic transport in \chem{PbF_2} and \chem{AgI}.\cite{CastiglioneEtAl_JPhysCondensMatter1999, CastiglioneAndMadden_JPhysCondensMatter2001, MorganAndMadden_JPhysCondensMat2012, MorganAndMadden_AgIDiffusionInSubmission}

\begin{figure*}[tb]
  \begin{center}
    \resizebox{17cm}{!}{\includegraphics*{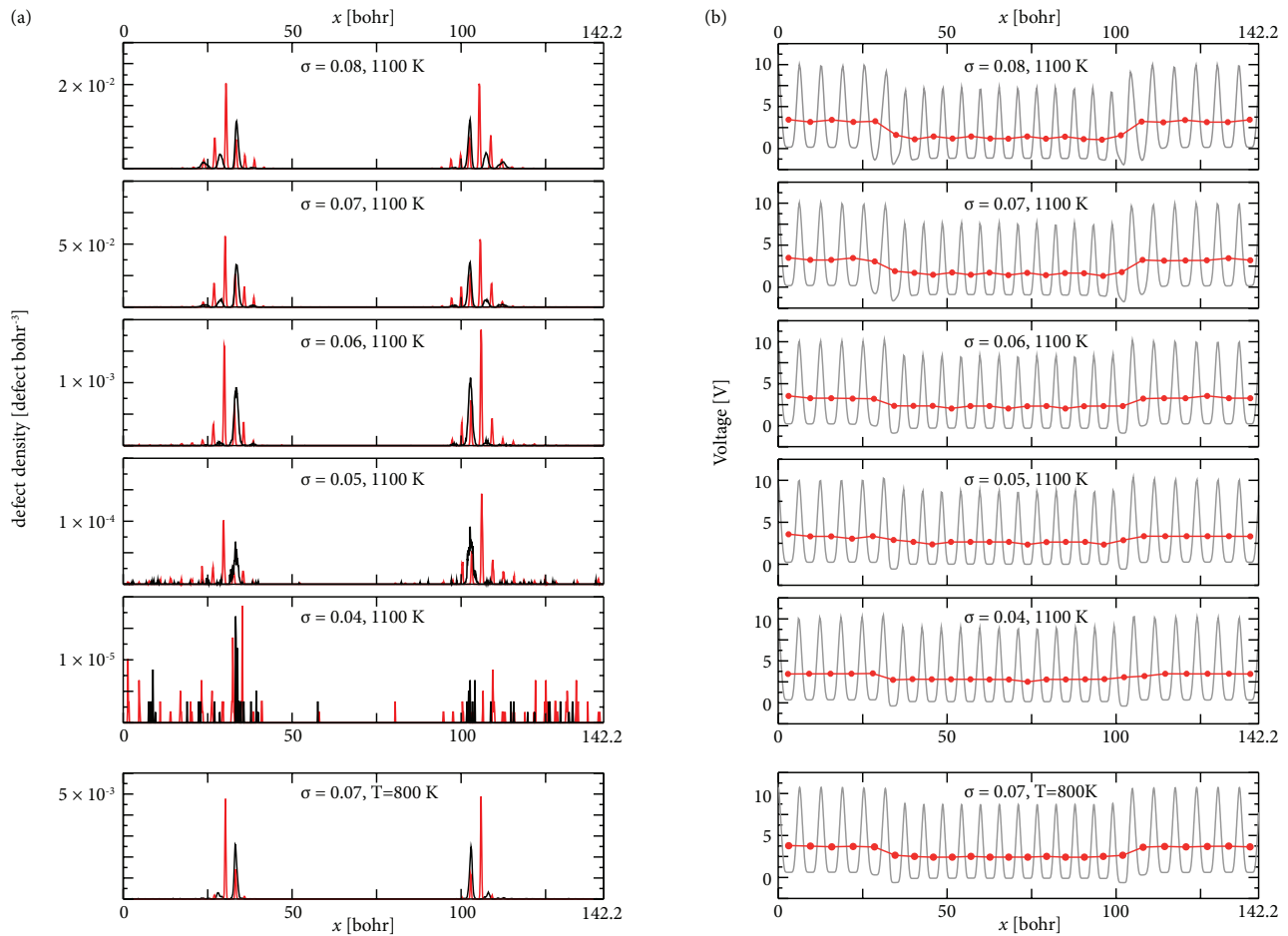}} %
    \caption{\label{fig:defect_poisson_1100K}(Color online) (a) Time averaged defect distributions from molecular dynamics simulations at $1100\,\mathrm{K}$. Fluoride interstitials are shown in black, vacancies are shown in red. (b) Poisson potentials calculated from molecular dynamics simulations at $1100\,\mathrm{K}$ for $\sigma=0.04$, $0.05$, $0.06$, $0.07$, and $0.08$. Grey lines: calculated potentials; black circles: average potentials for each Ca--F--F--Ca $(111)$ layer.}
  \end{center}
\end{figure*}

To directly compare defect distributions from our three-dimensional simulations to those predicted by the one-dimensional continuum mean-field model we project time-averaged interstitial- and vacancy-distributions onto the simulation cell $x$-coordinate (Fig.\ \ref{fig:defect_poisson_1100K}a); averaging out any temporal fluctuations and any defect structure in the perpendicular $y$ and $z$ directions. Calculated one-dimensional defect-density profiles for $\sigma=0.04$--$0.08$ at $T=1100\,\mathrm{K}$ reveal that defect densities are enhanced near the interface, with this effect increasing in magnitude and range with $\sigma_\mathrm{F}$. This is consistent with the enhanced fluoride ion mobilities in these same regions (Fig.\ \ref{fig:CaF2_heterostructure_diffusion_profile}). The defect enhancement extends only a few lattice spacings from the interface centre, which suggests that the continuum description may not be appropriate for this system. Interstitial and vacancy components of Frenkel pairs do not segregate to opposite sides of each interface, but instead numbers of interstitials and vacancies are increased relative to their bulk values on \emph{both} sides of the interfaces. These defect distributions are therefore qualitatively inconsistent with the picture of space charge formation predicted by the continuum mean-field model. Furthermore, at several $x$-positions both vacancy and interstitial numbers are enhanced; which is incompatible with the one-dimensional Poisson-Boltzmann equation. 

An alternative perspective of the defect distributions is offered by considering the mean electrostatic potential (Poisson potential) in each simulation, since this identifies regions of local non-charge neutrality. The mean electrostatic potential is calculated by integrating Poisson's equation, $\nabla^2\Phi=\frac{-\rho}{\epsilon_0}$, where $\rho$ is the mean charge density obtained from the simulations, and $\epsilon_0$ is the permittivity of free space. This potential is plotted in Fig.\ \ref{fig:defect_poisson_1100K}b for $\sigma=0.04$--$0.08$ and $T=1100\,\mathrm{K}$. The sharp oscillations in the calculated potential (grey lines) are due to the alternating F--Ca--F--F--Ca--F sequence of $(111)$ layers along the $x$ direction. An average potential can be calculated for each Ca--F--F--Ca $(111)$ layer by considering the sets of points between pairs of adjacent maxima in the calculated potential. In the bulk regions this average potential is flat, and is lower in region II than region I with an offset that increases with larger values of $\sigma_\mathrm{F}$. The voltage drop across the interfaces indicates a redistribution of ions, and is qualitatively similar to the behaviour predicted by the continuum model: the F$^-$ ions experience a high electrical potential in regions where the standard chemical potential is low (region II), and a low electrical potential where the standard chemical potential is high (region I). The small distances between vacancy and interstitial density peaks in the defect density profiles (Fig.\ \ref{fig:defect_poisson_1100K}a) indicate that defects close to the interface can be predominantly characterised as populations of associated Frenkel pairs. The observed voltage drop therefore cannot be explained by a segregation of interstitials and vacancies to form oppositely-charged atmospheres of defects on either side of the interface.

\subsection{Frenkel Polarisation and Concerted Diffusion}
\begin{figure}[tb]
  \begin{center}
    \resizebox{6cm}{!}{\includegraphics*{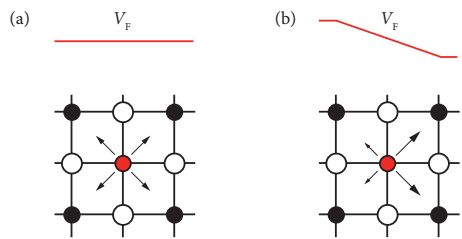}} %
    \caption{\label{fig:Frenkel_pair_schematic}(Color online) Schematic of the effect of variation in the background potential on Frenkel pair formation: (a) Where the potential gradient is zero, Frenkel pairs from with equal probability in all equivalent orientations. (b) Where the potential gradient is non-zero, specific orientations become favoured. A weighted average over all possible orientations can be expressed as a local dipolar field. In addition, the total probability of Frenkel pair formation is increased relative to the zero-field case.}
  \end{center}
\end{figure}

The electrostatic interactions of an associated Frenkel pair can be considered as those of a local dipole. In a bulk phase, Frenkel pairs form with a distribution of orientations that mirrors the underlying crystal symmetry. If a crystal structure has inversion symmetry, as for fluorite, the average local dipole of these Frenkel pairs is zero, and they do not contribute to the mean electrostatic potential (Fig.\ \ref{fig:Frenkel_pair_schematic}a). At a heterostructure interface the bulk crystal symmetry is broken, and Frenkel pairs may form with an asymmetric distribution of orientations, giving a net dipole (Fig.\ \ref{fig:Frenkel_pair_schematic}b). This ``Frenkel polarisation'' is an entirely dynamical process at low to moderate interfacial offsets of the fluoride standard chemical potential: for $\sigma\leq0.07$ quenching from the simulation trajectory to $T=0\,\mathrm{K}$ recovers the ideal fluorite lattice. For $\sigma\geq0.08$ these oriented Frenkel pairs are stabilised sufficiently by the external potential that they persist at $0\,\mathrm{K}$, indicating a defective ground state geometry. At $T>0\,\mathrm{K}$ the interface geometry can be thought of as dynamically sampling both perfect and Frenkel polarised structures, with the proportion of each depending on $\sigma_\mathrm{F}$ and temperature.

Dynamical formation of associated Frenkel pairs with preferred orientations explains the interfacial voltage drop without needing to invoke the formation of extended space charges. The voltage profile in each simulation is consistent with associated Frenkel pairs being stabilised at the interface with an orientation that places the fluoride interstitial in region II and the associated vacancy in region I, as would be expected from the independent defect chemical potentials (Fig.\ \ref{fig:CaF2_defect_energies}b). The drop in electrical potential going from region I to region II is complete within a couple of lattice spacings: outside of this region, and only a short distance from the interface, the lattice is on average locally charge neutral. The regions of enhanced defect numbers and increased local diffusion rates (Figs.\ \ref{fig:defect_poisson_1100K}a and \ref{fig:CaF2_heterostructure_diffusion_profile}a), however, are wider than these regions of non-zero potential gradient. These defects that lie outside of the region closest to the interface, therefore, do not, on average, contribute to a redistribution of charge perpendicular to the interfacial plane. The presence of regions of enhanced defect populations that are locally charge neutral allows us to infer details of the diffusion mechanism that follows the formation of associated Frenkel pairs at the interface.

Net ionic transport requires diffusion through the lattice of interstitials and vacancies generated as Frenkel pairs. Defect diffusion is typically characterised as a sequence of thermally activated hops giving intersitial or vacancy diffusion.\cite{Catlow_AnnRevMaterSci1986} If the diffusion of defects into regions away from the interface occurred by a series of independent single-ion steps, then the simulation trajectories would include configurations with excess interstitials or vacancies at these $x$ coordinates away from the interface. In this case the calculated electrostatic potential would have a non-zero gradient  across the whole of the region where we observe enhanced ionic diffusion. This is inconsistent with the simulated behaviour: the calculated electrostatic potential has zero gradient away from the interface, even in regions where the diffusion coefficient is locally enhanced, and therefore diffusion in these regions is not well described as a series of independent defect hops. An alternative mechanistic description, where diffusion is effected by a number of ions undergoing concerted motion, allows diffusion to occur without requiring movement of charge perpendicular to the interface, and is consistent with the broader widths of the calculated diffusion and defect concentration profiles compared to the calculated mean electrostatic profiles. 

\begin{figure}[tb]
  \begin{center}
    \resizebox{8cm}{!}{\includegraphics*{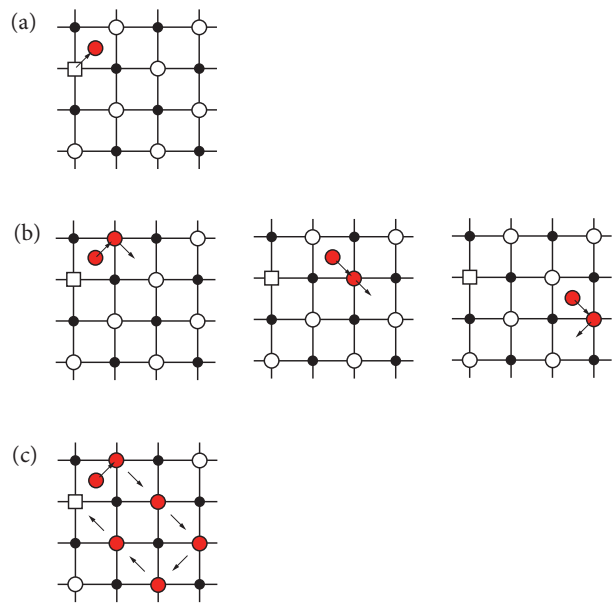}} %
    \caption{\label{fig:CaF2_concerted_diffusion}(Color online) (a) Diffusion is initiated by Frenkel pair formation at the interface. (b) Sequential interstitial hopping produces intermediate configurations with excess defects (locally non-charge-neutral at positions separated from the interface. (c) Concerted diffusion processes that start and end as associated Frenkel pairs at the interface give a electric field at positions separated from the interface.}
  \end{center}
\end{figure}

\subsection{Temperature Effects and the Debye Length}

The continuum mean-field model predicts that under Gouy-Chapman conditions the defect excess formed at a heterostructure interface decays to the bulk defect concentration on a length scale characterised by the Debye length, $\lambda_\mathrm{D}$; a characteristic screening length in each phase that depends on the dielectric constant and defect concentration in the bulk.\cite{Maier_ProgSolidStatChem1995, Maier_JPhysChemSol1985, Maier_BerBunsengesPhysChem1985,JamnikEtAl_SolStatIonics1995} Because defects are predicted to segregate into locally-non-stoichiometric space charge regions the mean electrostatic potential varies on the same length scale (Fig.\ \ref{fig:Space_Charge_Schematic}). 

The Debye length in solid electrolytes is given by
\begin{equation}
\label{eqn:debye_length}
  \lambda_\mathrm{D} = \left( \frac{\epsilon\epsilon_0kT}{2e^2c_\infty} \right)^\frac{1}{2},
\end{equation}
where $\epsilon$ is the relative dielectric constant, $\epsilon_0$ the absolute dielectric constant, $e$ the electron charge, and $c_\infty$ the bulk defect concentration. The defect distributions from our simulations differ qualitatively from those predicted by the continuum mean-field model, and we therefore would not expect Eqn.\ \ref{eqn:debye_length} to accurately describe the range of charge non-neutrality and defect disorder from the interface. Furthermore, the regions of  enhanced defect disorder and local non-charge-neutrality extend over only a few lattice spacings of the interface (Fig.\ \ref{fig:defect_poisson_1100K}), and at such small thicknesses the continuum model likely does not apply.\cite{JamnikEtAl_SolStatIonics1995} However, it is still instructive to compare values of $\lambda_\mathrm{D}$ obtained from Eqn.\ \ref{eqn:debye_length} with the spatial extent of the regions of local non-charge-neutrality  in the simulations. 

\begin{figure}[tb]
  \begin{center}
    \resizebox{5cm}{!}{\includegraphics*{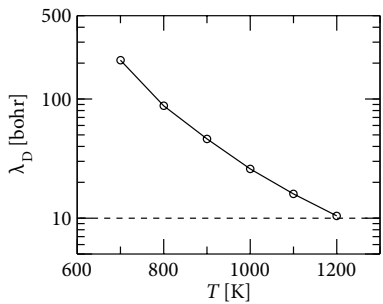}} %
    \caption{\label{fig:CaF2_debye_length}$\lambda_\mathrm{D}$ calculated according to Eqn.\ \ref{eqn:debye_length}, using $c_\infty$ calculated from bulk \chem{CaF_2} simulations. The dashed line gives an approximate half-width for the regions of defect enhancement at the simulated heterostructure interfaces.}
  \end{center}
\end{figure}

To calculate $\lambda_\mathrm{D}$ we use a value appropriate for this interatomic potential of $5.2$ for $\epsilon$ (compared to the experimental value of $\epsilon=6.47$)\cite{BinghamEtAl_JPhysCondensMat1999} and calculate $c_\infty$ from bulk \chem{CaF_2} simulations at the appropriate temperature, using the same criteria to identify defects as for the defect distribution plots. Because $\lambda_\mathrm{D}$ depends on the bulk defect concentration the calculated values have an approximate inverse exponential dependence on temperature (Fig.\ \ref{fig:CaF2_debye_length}). At $1100\,\mathrm{K}$, $\lambda_\mathrm{D}=16\,\mathrm{bohr}$ ($0.85\,\mathrm{nm}$), which is comparable to half the thickness of the regions of enhanced defect populations at this temperature ($\sim10\,\mathrm{bohr}$), although larger than half the width of the non-zero gradient regions in the mean electrostatic potential. If the distribution of defects at the heterostructure interface was accurately described by the continuum Poisson-Boltzmann equation then decreasing the temperature would be expected to \emph{increase} the width of regions of local non-charge-neutrality, as $\lambda_\mathrm{D}$ would increase. Our simulations, however, show different behaviour. At the lower temperature of $T=800\,\mathrm{K}$ ($\sigma_\mathrm{F}=0.07$) defect numbers are increased only immediately adjacent to the interfaces, and the mean electrostatic potential decays to bulk values over a \emph{smaller} distance than at $1100\,\mathrm{K}$ (Fig.\ \ref{fig:defect_poisson_1100K}). Both these measures of defect disorder follow the opposite trend to that predicted by the continuum mean field model. 

The reduction in width of the region of defect disorder with decreased temperature is consistent with the Frenkel polarisation mechanism. Associated Frenkel pairs are still formed preferentially at the interface, but in decreased concentrations, as seen from the reduced peak height in the defect densities. Reducing the temperature also decreases the probability of extended diffusion processes, which is consistent with a concerted diffusion process with a free energy barrier that increases with either the number of ions involved, or with the distance of the furthest ion from the interface. A reduction in temperature increases the probability that only short processes occur, restricting diffusive processes and associated defect disorder to the layers adjacent to the interface.

\section{Summary and Discussion}

To examine the role of crystal structure on the behaviour of mobile ions in heterostructured solid electrolytes we have constructed an atomistic model of a coherent ``\chem{CaF_2}/\chem{CaF_2}'' heterostructure, where a difference in the standard chemical potentials of the mobile fluoride ions across the interface is induced by an external potential. This is analogous to the construction used to describe intrinsic Gouy-Chapman conditions within the continuum mean-field model, which predicts segregation of defects into extended space charge regions on either side of the interfaces, with each space charge region described by a local excess of only a single defect type.

Molecular dynamics simulations of this coherent atomistic model show that increasing the offset in fluoride ion standard chemical potentials across the internal interfaces produces a large increase in fluoride ion diffusion coeffcients and net ionic conductivity, and strongly enhances fluoride ion mobilities and defect concentrations near the heterostructure interfaces. Concentrations of fluoride ion interstitials and vacancies are increased on \emph{both} sides of the interface, which is qualitatively inconsitent with the predictions of the continuum mean-field model. Instead of defects segregating into extended space charge regions with opposite formal charges, populations of associated Frenkel pairs with preferred orientations form at the interface. This ``Frenkel polarisation'' introduces an offset in electrostatic potential between the two regions, which contributes to restoring the equilibrium condition that the fluoride ion electrochemical potential is the same throughout the system. Local charge neutrality is restored within a few lattice spacings of the interface, at which  lengthscales a continuum model of defect behaviour is not expected to apply.\cite{JamnikEtAl_SolStatIonics1995} Considering together the calculated electrochemical potentials and defect profiles suggests that the enhanced diffusion at the interface is initiated by the formation of the associated Frenkel pairs at the interface, and proceeds by concerted diffusion processes that involve the motion of multiple fluoride ions.

The application of the continuum mean-field model follows from the assumptions that ionic defects in solid electrolytes can be considered as point charges in a continuum dielectic interacting only through a mean-field Coulomb term. This is similar to the textbook treatment of the distribution of electrons and holes at semiconductor $p$--$n$ junctions. We can speculate that the difference in behaviour between ionic and electronic defects is because ionic defects are highly localised charge centers that interact strongly with the lattice through both Coulombic and strain interactions. These interactions can cause defect positions to be strongly correlated, in addition to the direct defect--defect Coulomb correlations and excluded volume effects that would be present even in an isotropic system, and which give strong deviations from mean-field Gouy-Chapman behaviour in, for example, ionic liquids.\cite{Kornyshev_JPhysChemB2007} Electrons and holes, in contrast, are more delocalised as quantum particles, and interact relatively weakly with the lattice. The existence of strong correlations between defects explain the formation of associated Frenkel pairs, instead of segregation of interstitials and vacancies to \emph{opposite} sides of the interface predicted by the mean-field Space Charge Model, and is also consistent with concerted diffusion processes.

Our simulations also differ from the standard one-dimensional mean-field treatment in being three-dimensional. In a homogeneous phase, the Boltzmann term in the Poisson-Boltzmann equation provides a $1:1$ relationship between the concentration of a defect species at a point, $c_i$ and the local electrostatic potential, $\Phi$:
\begin{equation}
   c_i = c_\infty \exp\left(\frac{-q_i\Phi}{kT}\right).
\end{equation} 
For a non-zero electrostatic potential, each defect species is predicted to be either enhanced or depleted---which depends on the signs of the local electrostatic potential and the defect charge. The Poisson-Boltzmann equation therefore has no solutions with coincident concentration enhancements of oppositely charged defects. Describing a three-dimensional system with a one dimensional model implicitly assumes spatial homogeneity in the two perpendicular dimensions. A one-dimensional Poisson-Boltzmann model of a Frenkel-disordered crystal can not therefore have solutions with both defect species enhanced at the same $x$-position. Crystals are not spatially homogeneous, but periodically symmetric. It is therefore possible for defect distributions to be \emph{inhomogeneous} along $y$ and $z$ directions, as we find in our simulations (c.f.\ Fig.~\ref{fig:CaF2_heterostructure_diffusion_profile}b). If positions at different $y$ and $z$ coordinates but the same $x$ coordinate separately show enhanced vacancy and interstitial numbers, then projecting these defect populations onto one dimension will appear as coincident defect enhancements---as seen in Fig.~\ref{fig:defect_poisson_1100K}a.\footnote{Ions in a liquid electrolyte may move to any position, irregardless of their charge. In contrast, defects in a crystalline lattice are defined with reference to the parent crystal structure: interstitials exist only at interstitial sites, vacancies exist only at lattice sites. Because interstitial and vacancy sites are disjoint sets of mutually exclusive positions, it is not possible in three dimensions for interstitial and vacancy populations to overlap.}  It is therefore possible that a Poisson-Boltzmann model that accounts for crystalline symmetry both perpendicular and parallel to an interface may give qualitatively different predictions to an otherwise equivalent one-dimensional model, even if defect--defect correlations remain neglected.

In a heterostructure where both phases exhibit Frenkel disorder, and under intrinsic Gouy-Chapman conditions, all defects form thermally as components of Frenkel pairs. Within a crystalline lattice, these defects may interact strongly through both excluded volume and strain effects, and behaviours can arise that are qualitatively different to those predicted by the continuum mean-field model. In the model system considered here strong defect--defect interactions explain oscillatory defect profiles and enhanced ionic transport mediated by the concerted diffusion of multiple mobile ions. The simulation reported here considers intrinsic Gouy-Chapman conditions, and cannot be compared directly to heterostructured systems where extrinsic defects play a critical role in the defect chemistry (Mott-Schottky conditions), for example \chem{CaF_2}/\chem{BaF_2}.\cite{Jamnik_SolStatIonics2006,GuoEtAl_PhysRevB2007, GuoAndMaier_AdvFunctMater2009} In such systems, however, lattice effects and defect correlation may still be significant, and it might be necessary to account for these to obtain a full understanding of defect behaviour and associated ionic transport. More generally, the results presented here show that predictions from continuum mean-field models can differ qualitatively from equivalent systems when atomistic structure is taken into account. Such atomic structure effects may therefore also be important for understanding ``space-charge'' phenomena in other systems, such as metal/solid-electrolyte and intercalation-electrode/solid-electrolyte interfaces.

\section{Acknowledgement}
This work was supported by EPSRC Grant No. EP/H003819/1.

\bibliography{Bibliography}

\end{document}